\documentclass[referee]{aa}  % LaTeX A&A  Monotype Times Fonts
\usepackage{psfig}

\newcommand{\bcma}{$\beta$~CMa}
\newcommand{\acma}{$\alpha$~CMa}
\newcommand{\ecma}{$\epsilon$~CMa}

\newcommand{\cmd}{cm$^{-2}$}
\newcommand{\cmt}{cm$^{-3}$}
\newcommand{\kms}{km\,s$^{-1}$}

\newcommand{\cms} {cm$^{3}$\,s$^{-1}$} 

\def\lyb{Ly$\beta$}

\newcommand{\hei}  {\mbox{He\,{\sc i}}}
\newcommand{\hi}  {\mbox{H\,{\sc i}}}

\newcommand{\nitii}  {\mbox{N\,{\sc ii}}}
\newcommand{\niti} {\mbox{N\,{\sc i}}}

\newcommand{\oi} {\mbox{O\,{\sc i}}}
\newcommand{\ovi}  {\mbox{O\,{\sc vi}}}

\newcommand{\cii}  {\mbox{C\,{\sc ii}}}

\newcommand{\mgii} {\mbox{Mg\,{\sc ii}}}

\newcommand{\mgi}  {\mbox{Mg\,{\sc i}}}
\newcommand{\silii} {\mbox{Si\,{\sc ii}}}
\newcommand{\siliii}{\mbox{Si\,{\sc iii}}}

\newcommand{\sii}{\mbox{S\,{\sc ii}}}

\newcommand{\siii} {\mbox{S\,{\sc iii}}}

\newcommand{\feii} {\mbox{Fe\,{\sc ii}}}

\begin{document}

   \thesaurus{09         % A&A Section 8: ISM
              (09.01.1;  % abundances
               09.01.2;  % atoms
               09.02.1;  % bubbles
               09.03.1;  % clouds
               08.09.2;  % stars: \bcma
               13.21.3)  % UV : ISM
             }
   \title{Electron densities, temperatures and ionization rates in two
interstellar clouds in front of $\beta$ Canis Majoris, as revealed by UV
absorption lines observed with IMAPS}

   \author{Edward B. Jenkins$^{1}$, C\'ecile Gry$^{2,3}$, Olivier
Dupin$^{3}$}

   \offprints{C. Gry}

   \institute{$^{1}$ Princeton University Observatory,
              Princeton, NJ 08544-1001, USA
              (ebj@astro.princeton.edu)\\
              $^{2}$ ISO Data Center, ESA Astrophysics
              Division, PO box 50727, 28080 Madrid, Spain (present address)
              (cgry@iso.vilspa.esa.es)\\
              $^{3}$ Laboratoire d'Astronomie Spatiale,
              B.P.8,
              13376 Marseille cedex 12,
              France}

   \date{Received ; accepted }

   \authorrunning{Jenkins, Gry \& Dupin}
   \titlerunning{The Sight-line toward $\beta$~CMa}
   \maketitle

   \begin{abstract}
The spectrum of \bcma\ (Mirzam) between 1000 and 1200\AA\ was recorded
at a wavelength resolving power $\lambda/\Delta\lambda\sim 60\,000$ by
the Interstellar Medium Absorption Profile Spectrograph (IMAPS) during
its orbital flight on the ORFEUS-SPAS~II mission in 1996.  New
information about interstellar absorption lines of \cii, \cii*, \niti\
and \oi\ from this spectrum are combined with the HST results reported
by Dupin \& Gry (1998) to arrive at new conclusions about the physical
properties of the absorbing gas in front of \bcma.  For two prominent
velocity components centered at heliocentric velocities of +20.0 and
$+30.5\,$\kms, designated by Dupin \& Gry as Components C and D,
respectively, we use determinations of $N(\cii*)/N(\cii)$ and
$N(\mgi)/N(\mgii)$ to solve for temperatures and electron
densities. From our knowledge that oxygen and nitrogen have their ionizations
coupled to that of hydrogen through charge exchange reactions, we can
derive the hydrogen ionizations by comparing these elements to sulfur,
which is likely not to be depleted onto dust grains.  For Component~C with an
approximate column density of neutral and ionized hydrogen $N({\rm H}_{\rm
total})=6\times 10^{18}{\rm cm}^{-2}$,
we find that the neutral fraction $n(\hi)/n({\rm H_{\rm total}})=0.25$,
$400<T<6500\,$K, and
$0.08<n(e)<0.6\,$\cmt, while for Component~D with $N({\rm H}_{\rm
total})=1.2\times 10^{19}{\rm cm}^{-2}$, we arrive at
$n(\hi)/n({\rm H_{\rm total}})=0.035$, $8000<T<14\,000\,$K, and
$0.09<n(e)<0.2\,$\cmt.  The relatively large ionization fractions of
H can arise if the clouds are about 130~pc away from us, so that they
are exposed to the strong, ionizing radiation fields from $\epsilon$ and
\bcma.  The presence of \siliii\ indicates the existence of additional gas with
even higher levels of ionization.

\end{abstract}

\section{Introduction}
The line of sight toward the B1 II-III star \bcma\ in the direction 
($l_{II},b_{II}$)=($226\fdg,-14\fdg$) has raised interest since
its first study in the UV from {\it Copernicus} observations when Gry et al.
(1985) showed  that it contained very little neutral gas and that a great
majority
of the material on the line of sight was ionized.  Indeed for a distance now
set to
153 pc by Hipparcos measurements, Gry et al (1985) determined that $N$(\hi) was 
somewhere within the
range $1-2.2~10^{18}$\cmd\ based on the interstellar \lyb\ absorption
profile in the spectrum of \bcma\ recorded by {\it Copernicus}, and showed 
that the total (neutral and ionized) hydrogen column density was 
ten times higher using the combined
abundances of \sii\ and \siii\ compared to that of \hi\ multiplied by
the cosmic ratio of S to H. EUVE observations  
independently confirmed 
both findings. Cassinelli et al (1996) found 
N(\hi)=$2.0\pm0.2~10^{18}$\cmd\ by fitting the Lyman limit 
absorption
in the EUV spectrum of the star.
Evidence for ionization comes from the lower limits for the continuum
absorption by \hei\ in the spectrum of \bcma\ which indicate that
$N(\hei)>6.~10^{17}$\cmd\ (Aufdenberg, et al. 1999) or
$>1.4~10^{18}$\cmd\ (Cassinelli, et al. 1996), depending on assumptions
about the drop in stellar flux across the \hei\ ionization edge.  The
fact that $N(\hei)\gg 0.1N(\hi)$ indicates that a substantial fraction
of the hydrogen toward \bcma\ is ionized. 

The weak presence of \hi\ gas in this region of the sky is particularly 
evident in  the neighbouring
line of sight toward \ecma\ ($l_{II}=239\fdg8$, $b_{II}=-11\fdg3$) 
where the Lyman limit absorption is even lower 
(Cassinelli et al, 1995). Gry et al (1995) found a \hi\ column density 
upper limit of $5~10^{17}$ and showed furthermore that almost all the neutral
gas is very close to the Sun, distributed in two small components which
are also detected with similar column densities in the line of sight to 
Sirius 2.7 pc away (Lallement et al.,1994).
Because of the scarcity of \hi\ in this region, 
\ecma\ and \bcma\ are  the two strongest sources of
EUV radiation, and the two stars, especially \ecma, dominate
the H-ionization field close to the Sun (Vallerga \& Welsh 1995 ; Vallerga
1997). The conditions along these lines of sight, especially their ionization
structure, strongly influence the 
nature and ionization of the local interstellar matter. 
Yet even with this in mind, we find that the ionization of the local ISM is not
very well understood.  For instance, with the radiation from $\epsilon$ and \bcma\
plus other easily identifiable sources of ionizing radiation, it is difficult to
reconcile the fact that helium appears to be more ionized than hydrogen, as
revealed by evidence from the EUVE spectra of several white dwarfs (Dupuis et al.
1995).

High resolution observations with GHRS on board HST allowed Dupin and Gry 
(1998) to show that the bulk of the gas in the \bcma\ sight-line 
is distributed in two main
components separated by 10 km/s, which are warm, only slightly depleted
and both mostly ionized.
However the analysis of the ionization processes was hampered by the fact that
several species and especially the neutral species had  been observed
with a limited resolution of about 20~000 which did not allow 
for a clear separation of the two components. The relative distribution of 
matter in the
two components therefore had  uncertainties that allowed only lower limits
to  the ionization fractions to be derived.

Here we present new observations performed with the Interstellar Medium
Absorption Profile Spectrograph (IMAPS) (Jenkins, et al. 1996), with the
aim of gaining further insights on the ionization structure and various
physical processes in the local interstellar medium.  The higher
resolution of 60~000 provided by IMAPS allowed us to resolve the
absorptions of the two main components which in turn permitted the
derivations of ionization fractions in each cloud.  We also determined
the clouds' electron densities using the observation of \cii*, which
ultimately led to new insights on the ionization process.
\section{Observations and data reduction}
The observations were carried out by IMAPS when it was operated on the
ORFEUS-SPAS~II mission that flew in
late 1996 (Hurwitz, et al. 1998).  IMAPS is an objective-grating echelle
spectrograph that was designed to record the spectra of bright,
early-type stars over the wavelengths from $\sim950$~\AA\ to
$\sim1150$~\AA\ with a high spectral resolution. For more details on the
instrument see Jenkins et al. (1998, 1996).

The spectra were extracted from the echelle spectral images using
special procedures developed by one of us (EBJ) and his collaborators on
the IMAPS investigation team.  A zero point in the wavelength
calibration was made by measuring the \oi* and \oi** telluric lines
detected at 1040.943~\AA\ and 1041.688~\AA. Table~\ref{line_list}
presents the interstellar lines detected in the data, with wavelengths
and $f$-values taken from Morton (1991).
\begin{table}
\begin{center}
\caption{Interstellar absorption lines detected in the \bcma\ IMAPS
data.\label{line_list}}
\begin{tabular}{lcc}
\hline
\noalign{\smallskip}
 Element & $\lambda$ & f\\
\noalign{\smallskip}
\hline
\noalign{\smallskip}
\cii  & 1036.337 & 1.23\ $10^{-1}$\\
\noalign{\smallskip}
\cii* & 1037.018 & 1.23\ $10^{-1}$\\
\noalign{\smallskip}
\niti & 1134.165 & 1.34\ $10^{-2}$\\
\noalign{\smallskip}
      & 1134.415 & 2.68\ $10^{-2}$\\
\noalign{\smallskip}
      & 1134.980 & 4.02\ $10^{-2}$\\
\noalign{\smallskip}
\nitii & 1083.990 & 1.03\ $10^{-1}$\\
\noalign{\smallskip}
\oi   & 1039.230 & 9.20\ $10^{-3}$\\
\noalign{\smallskip}
\silii & 1190.416 & 2.50\ $10^{-1}$\\
\noalign{\smallskip}
      & 1193.290 & 4.99\ $10^{-1}$\\
\noalign{\smallskip}
\hline
\end{tabular}
\end{center}
\end{table}
\section{Analysis}
\begin{table*}
\begin{center}
\caption{Velocities and column densities for various species toward
\bcma.\label{results}}
\begin{tabular}{lccccc}
\hline
\noalign{\smallskip}
 Component & $v_\odot$ & $N$(\cii) & $N$(\cii*) & $N$(\niti) & $N$(\oi)
\\
& (\kms) & (\cmd) & (\cmd) & (\cmd) & (\cmd) \\
\noalign{\smallskip}
\hline
\noalign{\smallskip}
A\dotfill&3.0&$\sim 1.7~10^{13}$&$\leq 1.~10^{11}$&$\leq
1.~10^{11}$&$\leq 1.~10^{11}$ \\
\noalign{\smallskip}
B\dotfill&11.0&$\sim 2~10^{15}$&$\leq 1.~10^{11}$&$\leq 4.~10^{12}$&
$7.5\pm 2.5~10^{13}$ \\
\noalign{\smallskip}
C\dotfill&20.0&saturated&$2.4\pm 0.1~10^{13}$&$9.6\pm 0.4~10^{13}$&
$7.6\pm 0.4~10^{14}$ \\
\noalign{\smallskip}
D\dotfill&30.5&saturated&$1.2\pm 0.2~10^{13}$&$2.7\pm 0.3~10^{13}$&
$2.2\pm 0.2~10^{14}$ \\
\noalign{\smallskip}
\hline
\end{tabular}
\end{center}
\end{table*}

The structure of velocity components in the sight-line toward \bcma\ has
already been determined by the analysis of GHRS HST data (Dupin \& Gry
1998)~: four components were detected, two of which (designated as
Components C and D) were strongly dominant, and the remaining two (A and
B) showed up only for the strongest lines.  In the study conducted by
Dupin \& Gry (1998), only the lines between 1800~\AA\ and 3000~\AA\
(\feii, \mgii, \mgi\ and \silii\ lines) were observed at a resolution
$\lambda/\Delta\lambda\sim 80\,000$. Because of the failure of the
Echelle~A grating at the time of the observations, all the species that
had lines between 1150~\AA\ and 1800~\AA\ were observed at medium
resolution ($\sim20\,000$). With a resolution of $\sim60\,000$, the
IMAPS spectrograph now allows us to analyse lines of some important
species like \niti\ and \oi\ at a resolution comparable to that of the
GHRS Echelle data.  As a consequence, we can resolve partially clouds
C and D which are
separated by only $10\,$\kms.  To facilitate a comparison between the
IMAPS data and the HST data that were presented by Dupin \& Gry (1998),
we interpreted the column density vs. velocity profiles in terms of the
components defined by them.  However, the data are not precise enough to
rule out other, slightly different combinations of components that could
produce fits that are just as acceptable.

The relative velocities between the four clouds are precisely known from
the Dupin \& Gry (1998) study.  The heliocentric radial velocities found
in our study are larger than those of Dupin \& Gry (1998) by about
$5.5\,$\kms.  This difference is larger than the stated error for the
GHRS velocity zero points, but we favor the velocities from the IMAPS
data because we could use the telluric absorption lines as a reference. 
The velocity of \cii* is especially good since it is very close to the
\oi**\ line at 1041.688\AA\ in the spectral format of the IMAPS echelle
and cross disperser gratings. The velocity dispersions $b$ for \cii,
\niti\ and \oi\ in Components C and D are not well determined, since
some blurring of the lines is caused by the instrument.  However, they
should be constrained by $b$(\mgii) on the low end and the measured
widths on the upper end.  Thus, the principal free parameters are the column
densities of various species in cloud C and cloud D.  For the strong
lines of \niti\ and \oi\ we see some evidence for the presence of
Component B, and we can place upper limits for the strength of A.

The principal source of uncertainty for the column densities is the
determination of the background intensity caused by scattered light from
the echelle and cross disperser gratings in IMAPS.  As is evident from
Jenkins, et al. (1996 -- see their Fig.~12) there is some overlap of
energy from adjacent echelle orders, precluding the use of the
interorder regions for measuring the background levels.  We could
determine the overall behavior of the scattered light by examining the
trends for the bottoms of strong stellar absorption features.  For the
background level below the line of \cii* at 1037.018\AA,
we could use the nearby, saturated absorption from interstellar \cii\ at
1036.337\AA.

\begin{figure*}
\centerline{\psfig{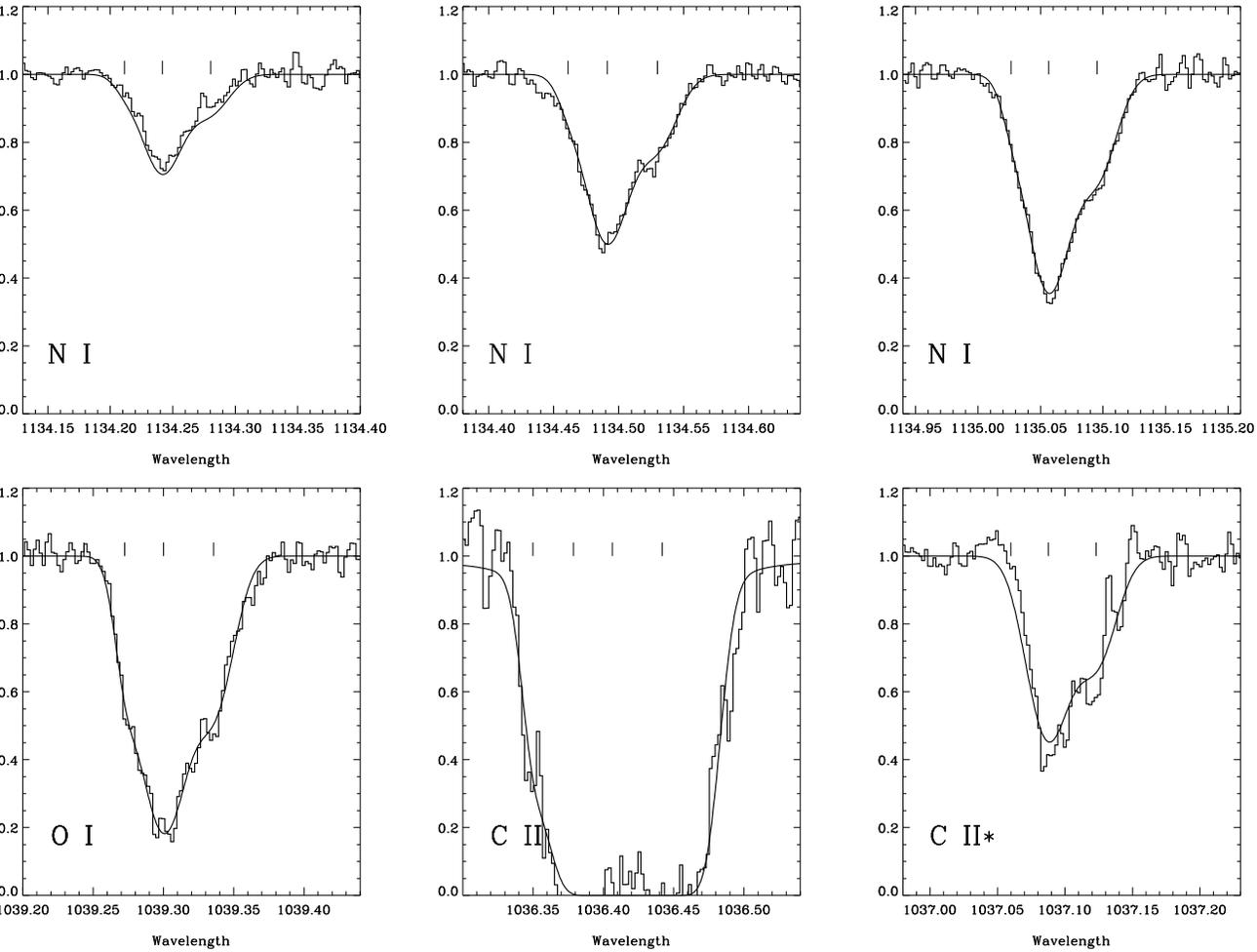}}
\caption{Interstellar lines observed with IMAPS. The histogram-style
tracings represent the recorded data after normalization to the stellar
continuum, and the smooth lines show the fits 
obtained from our analysis.  The tick marks above the absorption
features denote (left to right) the positions of the velocity Components
B, C and D identified in the text, except for \cii\ which shows
Component A (far left) in addition to the others.}\label{IMAPS_spec}
\end{figure*}

Fig.~\ref{IMAPS_spec} shows the features recorded by IMAPS after they were
normalized to the stellar continuum.  Overplotted on these data are the predicted
absorption profiles for the column densities and central velocities
given in Table~\ref{results}.  Selected results for Components C and D
derived by Dupin \& Gry (1998) are listed in Table~\ref{DG}, as an aid
to following the interpretations that we present in
\S\ref{interpretation}.  We made no attempt to measure the \silii\ lines
listed in Table~\ref{line_list}, since they were saturated and not
nearly as useful as the transition at 1808~\AA\ observed by Dupin \& Gry
(1998).  Conversely, our \niti\ lines are not saturated, but the weakest
member of the 1200~\AA\ multiplet observed by Dupin \& Gry had a central
optical depth $\tau_0=1.7$ for Component~C [derived from our $N$(\niti)
and their $b$ value].  The lack of saturation and higher resolution for
the IMAPS recordings of \niti\ absorption lead to our preferring the
values derived here to those of Dupin \& Gry, even though there is some
uncertainty in the IMAPS background level.

The \oi\ transition at 1302~\AA\ observed by Dupin \& Gry was hopelessly
saturated, which led to their only being able to express a lower limit
for $N$(\oi).  Fig.~\ref{IMAPS_spec} shows that the IMAPS recording of
the much weaker transition at 1039~\AA\ shows a fairly strong saturation.
If there are
narrow, unresolved subcomponents within Component~C that are far more
saturated than what we see in the apparent residual intensity of the
line, our listed column density may be below the true value.  In
principle, one could sense the presence of such components by observing
that the \niti\ absorptions do not grow as fast as their increases in
transition strengths.  However the lines that we observed here are too
weak to show this effect well.  Nevertheless, it is interesting to note
that by observing the strong multiplet of \niti\ at 1200~\AA\ Dupin \&
Gry (1998) found $N(\niti)=8.1\pm0.8~10^{13}$\cmd\ for Component~C 
(including the hidden local clouds)\footnote{Note that in
Dupin \& Gry Table 2, the listed column densities for Component~C
had a contribution from the LIC and the other local cloud  detected 
in both \acma\ and \ecma\ sightlines subtracted off. This contribution
is negligible for all species but \oi\ and \niti\ for which it represents
about 25\% of the total absorption in Component~C.},
which is only 0.84 times the value that we obtained (see
Table~\ref{results}). (Our and their results for Component~D agree
however.)  Thus, when lines reach a strength similar to that of the \oi\
line at 1039~\AA, the column densities might be understated by a factor
of about 0.84.

\begin{table}
\begin{center}
\caption{Selected results from Dupin \& Gry's (1998) study of
\bcma.\label{DG}}
\begin{tabular}{lccccc}
\hline
\noalign{\smallskip}
 Species & Comp. C & Comp. D \\
\noalign{\smallskip}
\hline
\noalign{\smallskip}
$N$(\sii) (\cmd)&$1.1\pm 0.4~10^{14}$&$2.3\pm 0.5~10^{14}$ \\
\noalign{\smallskip}
$N$(\silii)\dotfill&$5.8\pm 0.5~10^{13}$&$7.5\pm 0.3~10^{13}$ \\
\noalign{\smallskip}
$N$(\siliii)\dotfill&$\leq 5~10^{12}$&$(1.5-10)10^{14}$ \\
\noalign{\smallskip}
$N$(\mgi)\dotfill&$8.0\pm 0.6~10^{10}$&$2.5\pm 0.3~10^{11}$ \\
\noalign{\smallskip}
$N$(\mgii)\dotfill&$3.1\pm 0.6~10^{13}$&$5.9\pm 0.8~10^{13}$ \\
\noalign{\smallskip}
$N$(\feii)\dotfill&$1.2\pm 0.1~10^{13}$&$1.0\pm 0.1~10^{13}$ \\
\noalign{\smallskip}
\hline
\end{tabular}
\end{center}
\end{table}
\section{Interpretation}\label{interpretation}
\subsection{Radial velocities}\label{velocities}

As noted earlier, there is a systematic offset between the velocities
reported here and those given by Dupin \& Gry (1988).  Our new velocity
for Component~C at $20\,$\kms\ makes it consistent with that of gas in
the Local Interstellar Cloud (LIC), moving at a velocity of $26\,$\kms\
towards the direction $l=186^\circ$ and $b=-16^\circ$ (Lallement \&
Bertin 1992), whereas Dupin \& Gry felt that the LIC was buried in the
gap between Components C and D.  However, there are two important
problems for  identifying the origin of Component~C with the LIC. 
First, the column densities of \mgii\ and \feii\ in Component~C (see
Table~\ref{DG}) are considerably larger than
$N(\mgii)=1.65~10^{12}$\cmd\ and $N(\feii)=8.5~10^{11}$\cmd\ found in the LIC
component at $18.8\,$\kms\ in the spectrum of \acma\ (Lallement, et al.
1994) which is only $5.5^\circ$ away in the sky, and than 
$N(\mgii)=3.0~10^{12}$\cmd\ and $N(\feii)=1.35~10^{12}$\cmd\ found in the LIC
component at $17\,$\kms\ in the spectrum of \ecma\ (Gry, et al.
1995), $17^\circ$ away in the sky.  Also,
$N(\hi)=5~10^{17}$\cmd\ in front of \acma\ (Holberg, et al. 1998) is
only about a third as large as the \hi\ column density that we identify
with Component~C [$N(\hi)=1.6~10^{18}$\cmd, taken from $N$(\niti) and
$N$(\oi) after correcting for depletion -- see \S\ref{ionization}
below.]  Even though \acma\ is only 2.6~pc away, it seems to lie beyond
the boundary of the LIC because Lallement, et al. (1994) detected
another (somewhat weaker) component at $v=13\,$\kms.  Second, we show in
\S\ref{ionizing} that the ionizing radiation field that is needed to
sustain the ionization is considerably higher than that found for the
local vicinity.  Thus, the agreement of velocity of Component~C with
that of the LIC is probably just coincidental. 
\subsection{Column densities, depletions, and ionization
fractions}\label{ionization}

We define a depletion factor $\delta(X)$ of an element $X$ in terms of
its column density $N(X)$, relative to that of hydrogen $N({\rm H})$, by
the expression
\begin{equation}\label{deltaX}
\delta(X)=\left( {N(X)\over N({\rm H})}\right)~\left({{\rm H}\over
X}\right)_{\rm cosmic}
\end{equation}
where $({\rm H}/X)_{\rm cosmic}$ is the cosmic abundance
ratio.\footnote{Throughout this paper, we use the cosmic abundances of
Anders \& Grevesse (1989)}  For such a depletion to be meaningful in any
given circumstance, one must account for differences in the observed
ionization stage(s) of element $X$ relative to that adopted for H
(either \hi\ or H$_{\rm total}$).  We address this consideration in the
cases that follow, starting with the simplest ones involving the
elements nitrogen, oxygen and sulphur.

The relative ionizations of \niti\ and \oi\ are closely coupled to that
of \hi\ through the large charge exchange reaction rates that arise from
their nearly equal ionization potentials (Field \& Steigman 1971; Butler
\& Dalgarno 1979).  In the diffuse interstellar medium, these two
elements are only mildly depleted below their cosmic abundances relative
to hydrogen (Hibbert, et al. 1985; Cardelli, et al. 1991a,b; Meyer, et
al. 1994, 1997, 1998), so they can serve as reasonably good indicators
for \hi\ in a partially ionized region.  Sulphur is another element that
has little or no depletion onto dust grains, but its second ionization
potential is very high (23~eV).  If the photoionization rate arising
from photons with $E>23$~eV is not very large, we expect that \sii\
should be a good indicator of the total hydrogen in the line of sight,
both in the neutral and ionized forms.  In essence, for N, O and S
we can make use of our general understanding of their depletions to arrive at
conclusions on the relative ionization of hydrogen.

With our ability to identify how much of the absorptions by \niti\ and
\oi\ can be assigned to Components C and D, we can differentiate
between the relative
fractions of H in neutral form in each case, after making comparisons with
Dupin \& Gry's (1998) values for \sii\ in these same components.  In
doing this, we invoke two assumptions about the depletions: (1) the
depletions of N and O onto dust grains are the same for Components C and
D, and (2) S is undepleted.  Assumption (1) seems reasonable in the
light of evidence presented by Meyer, et al. (1997, 1998) and (2) seems
justified from the evidence summarized by Savage \& Sembach (1996) and
Fitzpatrick \& Spitzer (1997).  If either of these assumptions is not
quite correct, small errors in our conclusions may arise.  However, they
are probably not much worse than the uncertainty in identifying the
relative contributions of Components C and D in the low resolution
spectra of the \sii\ features.

For nitrogen, we obtain a general depletion factor $\delta({\rm N})=0.55$ through
Eq.~\ref{deltaX} by taking $N(\niti)_{\rm C+D}$, dividing it by
$N(\hi)=2.~10^{18}$\cmd, and then
multiplying the result by the cosmic ratio of H to N.  The same analysis
for O leads to $\delta({\rm O})=0.58$.  These values, incidentally, are
not much different from those shown by Meyer, et al. (1997, 1998).  We
now estimate the hydrogen neutral fractions in each component from the
expression (for the case of \niti)
\begin{equation}\label{neut_frac}
n(\hi)/n({\rm H_{\rm total}})=\left({{\rm S}\over {\rm N}}\right)_{\rm
cosmic}{N(\niti)\over
\delta({\rm N})N(\sii)}
\end{equation}
and likewise for \oi.  For both \niti\ and \oi, we obtain $n(\hi)/n({\rm
H_{\rm total}})=0.25$ for Component~C and 0.035 for Component~D.  

The high column density of \siliii\ identified with Component~D by Dupin
\& Gry (1998) presents a special problem (see Table~\ref{DG}).  The rate
constant for charge exchange between doubly ionized Si and neutral H is
about $3.0~10^{-9}$\cms\ at the temperatures of interest to us (Gargaud,
et al. 1982).  From the electron density derived later in
\S\ref{CII_exc} and $n(\hi)/n({\rm H_{\rm total}})$ for Component~D, we
find that $n(\hi)\sim 0.005\,$\cmt\ (for an applicable temperature that
we derive in \S\ref{Mg_ratio}).  This density times the charge exchange
rate constant is considerably larger than the expected ionization rate
$\Gamma(\silii)=5.~10^{-14}{\rm s}^{-1}$ that we infer from the hydrogen
ionization and the shapes of the EUV spectra of $\epsilon$ and \bcma\
(Vallerga \& Welsh 1995; Aufdenberg, et al. 1999) which should dominate
the photoionizing radiation field.  It is hard to reconcile this
inequality with Dupin \& Gry's finding that $N(\siliii)/N(\silii)\geq 2$
for Component~D.  We propose that this contradiction can be overcome by
having the \siliii\ exclusively within a different part of the region, perhaps one
that is much closer to some ionizing source or, alternatively, within the cloud's
conduction/evaporation front at an interface with gaseous material at a much
higher temperature.  There is empirical evidence that the association of \siliii\
with lower ionization stages is not unusual.  For instance, Cowie, et al. (1979)
found that the velocity endpoints of \ovi, \siliii, and \nitii\ features were
mutually correlated, and they used this evidence to suggest that the interfaces
between cool and hot gas were conspicuous in the interstellar absorption lines.

In sections that follow, we will ignore the existence of the
\siliii -bearing region and propose that the contradiction that we have
noted justifies our regarding it as unrelated to the gas that holds most of the
lower ionization states.
\subsection{Electron densities from $N$(\cii*)/$N$(\cii)}\label{CII_exc}
The relative populations of the fine-structure states of \cii\ are
governed by the balance between collisions and the radiative decay of
the upper level.  If electrons are the dominant projectiles for
excitation and de-excitation, the rate coefficient for de-excitations is
\begin{equation}\label{downward}
\gamma_{2,1}={8.63~10^{-6}\Omega_{1,2}\over g_2T^{0.5}}{\rm cm}^3{\rm
s}^{-1}
\end{equation}
(Spitzer 1978, p. 73), with the reverse rate given by detailed
balancing, $\gamma_{1,2}=(g_2/g_1)\exp(-E_{1,2}/kT)\gamma_{2,1}$.  The
statistical weights of the levels are $g_1=2$ and $g_2=4$, and the
temperature equivalent for the difference in energy levels
$E_{1,2}/k=94.9$K.  Thus we find that the condition for equilibrium,
\begin{equation}\label{equilib}
n(e)\gamma_{1,2}n(\cii)=[n(e)\gamma_{2,1}+A_{2,1}]n(\cii*)
\end{equation}
will lead to an equation for the electron density
\begin{equation}\label{n(e)}
n(e)={g_2A_{2,1}T^{0.5}\left[{n(\cii*)\over n(\cii)}\right]\over
8.63~10^{-6}\Omega_{1,2}\left\{ \left({g_2\over g_1}\right) \exp\left(
{-E_{1,2}\over kT}\right) - \left[{n(\cii*)\over n(\cii)}\right]\right\}
}
\end{equation}
where the radiative decay probability for the upper level is
$A_{2,1}=2.29~10^{-6}{\rm s}^{-1}$ (Nussbaumer \& Storey 1981), and the
collision strength $\Omega_{1,2}=2.81$ (Hayes \& Nussbaumer 1984).
Optical pumping of the \cii\ fine structure levels is unlikely to happen
under the circumstances where there is a very high optical depth in the
line (Sarazin, et al. 1979).  Also, the density of pumping radiation
must be very large -- even larger than the elevated levels considered
later in \S\ref{ionizing}.  Finally, unacceptably high values for
$n(\hi)$ would be needed for collisions by neutrals to have any
importance (Keenan, et al. 1986). 

As is clear from Fig.~\ref{IMAPS_spec}, the \cii\ feature at 1036~\AA\
is far too saturated to allow a determination of $N(\cii)$ for either
Components C or D.  Thus, our determination of the column densities must
be indirect.  A good surrogate for carbon is sulphur.  Our repeat of
calculations of the type done by Sofia \& Jenkins (1998) indicate that
in partially ionized gases these two elements have about the same
fraction of atoms elevated to higher (unseen) stages of ionization for a
wide range of conditions.  The depletion of carbon atoms in dense clouds
is typically $\delta({\rm C})=0.39$ (Cardelli, et al. 1993, 1996; Sofia,
et al. 1998), and, learning from the example of $\tau$~CMa (Sofia, et
al. 1997), this level of depletion seems to hold also for low density
lines of sight.  On the assumption that carbon toward \bcma\ is depleted
by this factor and there is no depletion of sulphur, we can arrive at
$N(\cii)$ from the product $({\rm C/S})_{\rm cosmic} \delta({\rm
C})N(\sii)$.  Doing so gives us the values
$N(\cii)=8.3~10^{14}$\cmd\ for Component~C and $1.7~10^{15}$\cmd\ for
Component~D, leading to $N(\cii*)/N(\cii)=0.029^{+0.020}_{-0.012}$ and
$0.0071^{+0.0034}_{-0.0023}$ for Components C and D, respectively.

Before we can derive values for $n(e)$ from Eq.~\ref{n(e)}, we must
determine $T$.  To do this, we rely on another method for measuring
$n(e)$, but one that has a different temperature dependence.  The
balance between the ionization of \mgi\ and the recombination of \mgii\
is a fundamentally different process from that which governs the
fine-structure equilibrium of \cii, but both are driven by the value for
$n(e)$.  In the next section (\S\ref{Mg_ratio}), we shall make use of
the difference to constrain other free parameter, the temperature $T$.
\subsection{Electron densities from \mgi/\mgii}\label{Mg_ratio}
The equation for the equilibrium of the lowest 2 ionization levels of Mg
is given by
\begin{eqnarray}\label{mg_equilib}
\big[\Gamma(\mgi)+C(\mgii)n(H^+)\big]n(\mgi)=&& \nonumber\\
\alpha(\mgi)n(e)n(\mgii)&&
\end{eqnarray}
For the charge exchange rate $C(\mgii)$ that applies to the reaction
\mgi~+~H$^+\rightarrow$ \mgii~+~H, we
used the analytical approximation
$C(\mgii)=1.74~10^{-9}\exp(-2.21~10^4/T)$ derived by Allan, et al. (1988). 
Collisional ionization of \mgi\ is negligible at the temperatures of
interest in this study.  We computed $\Gamma(\mgi)$ at the position of
the Sun using mean of the stellar radiation fields estimated by Jura
(1974) $\lambda F_\lambda=3.0~10^{-3}{\rm erg~cm}^{-2}{\rm s}^{-1}$ and
Mathis, et al. (1983) $\lambda F_\lambda=2.5~10^{-3}{\rm
erg~cm}^{-2}{\rm s}^{-1}$, multiplied by the photoionization cross
section given by Verner, et al. (1996), to arrive at
$\Gamma(\mgi)=6.1~10^{-11}{\rm s}^{-1}$.  (However, this number
increases to $9.5~10^{-11}{\rm s}^{-1}$ when we consider locations
closer to $\epsilon$ and \bcma\ at a later stage of the analysis.)  For
$\alpha(\mgi)$ we used the radiative and dielectronic recombination
rates given by Shull \& van Steenberg (1982), supplemented by the
additional contributions from low-lying resonance states computed by
Nussbaumer \& Storey (1986).

It is immediately evident that the Echelle~B observations of \mgi\ and
\mgii\ by Dupin \& Gry (1998) both show an excellent signal-to-noise
ratio, a smooth stellar continuum flux, and absorption features that are
reasonably well resolved.  Nevertheless, there is a difference between the two. 
On the one hand, the 2853~\AA\ feature of \mgi\ represents an absorption that
is unsaturated and thus straightforward to analyze.  On the other hand,
the two features of \mgii\ (2796 and 2803~\AA) were  heavily saturated,
which leads to a large uncertainty in $N(\mgii)$.  Even though Dupin \&
Gry could use the well-determined $b$ value for \mgi\ to help in the
interpretation of the \mgii\ features, the error in $N$(\mgii) is still
significant 
(see Table~\ref{DG}).  We will use here a less direct method of deriving
$N$(\mgii) from similar but more accurately determined species, along
with an application of some empirical relationships for interstellar
depletions, that will ultimately lead to a completely independent estimate for
$N$(\mgii).

As with C and S discussed in \S\ref{CII_exc}, Mg and Si are a good pair
of elements that have very similar photoionization properties (although
the rate coefficients for charge exchange of their doubly ionized forms
with neutral hydrogen are very different -- see below). Fitzpatrick
(1997) shows a general relationship between the logarithms
of the interstellar depletions
of Mg and Si.  If we assume that S is generally undepleted, we can
substitute it for H and then derive the quantity $\log\delta ({\rm
Mg})$ from our observed $\log\delta({\rm Si})$ and his best fit to
the trend $\log\delta({\rm Mg})=0.82 \log\delta({\rm Si})-0.17$~dex. 
When we do this, we find that $\log\delta({\rm Mg})=-0.63$~dex for
Component~C and $-0.80$~dex for Component~D.  Products of these derived
$\delta$'s, the \sii\ column densities, and the cosmic values for Mg
relative to S give $N(\mgii)=5.4~10^{13}$\cmd\ and $7.6~10^{13}$\cmd\
for Components C and D, respectively.  A moderate adjustment in
$N(\mgii)$ for Component~D must be made to account for the fact that
about 24\% of the Mg is probably in the doubly ionized form, while the
fractional amount of \silii\ is much lower because it has a
significantly larger rate coefficient for recombining via charge
exchange with neutral hydrogen.  This conclusion was derived
quantitatively after making preliminary calculations of the ionization
balance, as described later in \S\ref{ionizing}.  No such correction is
needed for Component~C. The estimate we made here for $N(\mgii)$ for
Component~C is somewhat larger than the one given in Table~\ref{DG} --
however is marginally consistent with it when the error of 0.16 dex attached 
to our present method is considered-- 
while our (adjusted) value of $N(\mgii)=5.8~10^{13}$\cmd\ is remarkably
close to the value derived by Dupin \& Gry.   Finally, we arrive at
$N(\mgi)/N(\mgii)$ equal to $0.0015^{+0.0007}_{-0.0005}$ for Component~C
and $0.0043^{+0.0020}_{-0.0013}$ for Component~D.  Note that while Dupin
\& Gry's values for $N$(\sii) have large estimated relative errors
(36\%, 22\%), such errors have a very weak influence in the conclusions
because the slope of the empirical relationship between $\log\delta({\rm
Mg})$ and $\log\delta({\rm Si})$ is not much different than 1.0.  The same
applies to errors in the assumption that $\log\delta({\rm S})=0.0$~dex.

As a check that the depletions toward \bcma\ are not out of the
ordinary, we carried out a similar exercise, comparing the measured
$\log\delta({\rm Fe})=(-1.20,~-1.60)$~dex and their respective
$\log\delta({\rm Si})=(-0.56,~-0.77)$~dex for consistency with the trend
shown by Fitzpatrick (1996).  Both cases fell within the  body of
points that defined this trend.

\begin{figure}
\centerline{\psfig{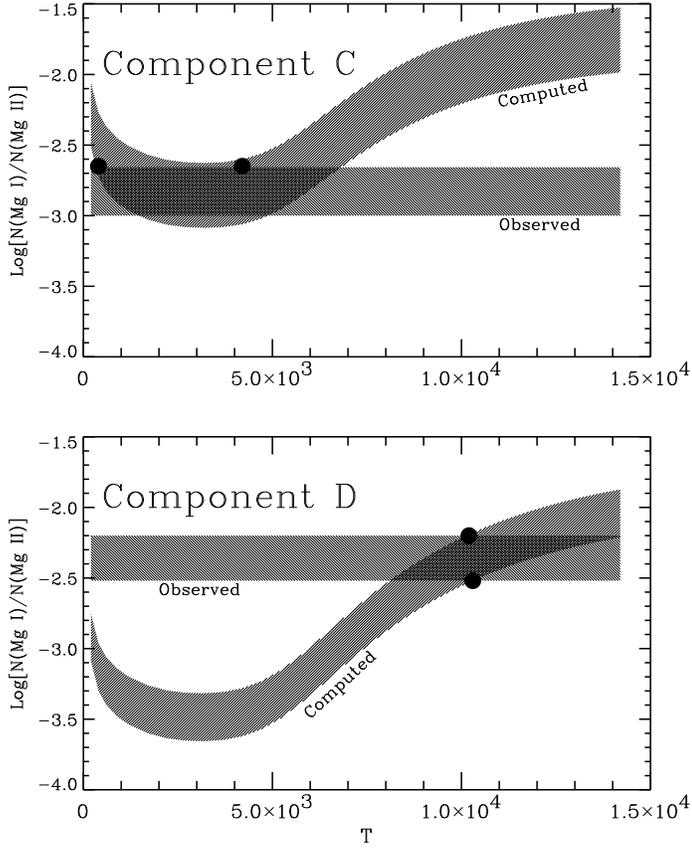}}
\caption{Predictions for the logarithms of $N(\mgi)/N(\mgii)$ as a
function of temperature $T$ (curved, shaded bands), computed from
Eq.~\protect\ref{mg_equilib} using electron densities $n(e)$ derived
from the measurements of $N(\cii*)/N(\cii)$ applied to
Eq.~\protect\ref{n(e)}.  The widths of these bands indicate the ranges
that are allowed by our estimated errors.  The shaded, straight,
horizontal bands indicate the observed values (and their uncertainties)
for $\log[N(\mgi)/N(\mgii)]$.  In each case, the overlaps of the
straight and curved bands indicate the temperature ranges that are
consistent with both the $N(\cii*)/N(\cii)$ and $N(\mgi)/N(\mgii)$
results.  Positions on the diagram that represent the worst extremes in
$n(e)$ are indicated by solid dots; see the text for
details.}\label{Mg_T}
\end{figure}

The curved bands in Fig.~\ref{Mg_T} show,
as a function of temperature $T$, the expected
outcomes for $\log N(\mgi)-\log N(\mgii)$ found by solving the ionization
equilibrium equation (Eq.~\ref{mg_equilib}) with the values of $n(e)$
obtained from Eq.~\ref{n(e)} and the measured values of
$N$(\cii*)/$N$(\cii) given in \S\ref{CII_exc}.  From the intersections
of these curves with the logarithms of the observed $N(\mgi)/N(\mgii)$
presented above (straight, horizontal bands in the figure),
we find that $400<T<6500\,$K for Component~C and
$8000<T<14\,000\,$K for Component~D.  To compute the upper limits for
$n(e)$, we evaluate Eq.~\ref{n(e)} with the upper limit for
$N(\cii*)/N(\cii)$ at the higest temperature that is consistent with
this upper limit ({\it not} the highest temperature allowed in general).
For the lower limits, we take the lowest permissible $N(\cii*)/N(\cii)$
and apply it to Eq.~\ref{n(e)} with the lowest temperature.  These
conditions for the limits are shown by solid dots in Fig.~\ref{Mg_T}. 
For Component~C we find that $0.08<n(e)<0.6\,$\cmt, and for
Component~D we arrive at $0.09<n(e)<0.2\,$\cmt. (While our lower limit
for $n(e)$ for Component~C is formally allowed, the real value of $n(e)$
is probably much closer to the upper limit because the temperature of
the gas is probably much higher than $400\,$K.)  These represent the
worst possible extremes in $n(e)$ permitted by the data.
\subsection{Ionizing radiation}\label{ionizing}
Shortward of the Lyman limit, the general ionizing radiation field near
the Sun is dominated by the output of $\epsilon$ and \bcma\,
supplemented by many nearby white dwarf stars, cataclysmic variables,
and the line emission from some late-type stars (Vallerga 1997).  In
addition to stellar sources, some diffuse energetic radiation from the
hot gas that surrounds the LIC may contribute additional ionization
(Cheng \& Bruhweiler 1990).  The intensity of this diffuse radiation is
rather uncertain: Jelinsky, et al. (1995) have observations that place a
meaningful upper limit for this radiation if it produces spectral
features in agreement with some specific predictions of hot-gas emission models.  

If we calculate the hydrogen ionization produced by Vallerga's (1997)
composite EUV stellar radiation field, supplemented by Cheng \&
Bruhweiler's (1990) hot gas radiation attenuated by absorption from a
hydrogen column $N(\hi)=2~10^{17}$\cmd, we find that the neutral
fraction of hydrogen is 0.89 and 0.47, for Components C and D,
respectively, if we require that $n(e)=0.31\,$\cmt\ and $0.13\,$\cmt\ as
we found from the analysis of $N(\cii*)/N(\cii)$ and $N(\mgi)/N(\mgii)$
in \S\S~\ref{CII_exc} and \ref{Mg_ratio}.  This is clearly inconsistent
with our finding $N(\hi)/N({\rm H_{\rm total}})=0.25$ and 0.035 for the
two components in \S\ref{ionization}.  The model calculations followed
the scheme outlined by Sofia \& Jenkins (1998) that included the
ionization of helium and allowed for the extra electrons coming from He. 
(In contrast to the LIC, where helium ionization is important, the
models indicated that the high values of $n(e)$ effectively suppressed
the ionization of helium.)

To overcome the fact that the predicted ionization of H is less than
observed, we must find additional sources of ionization.  One
possibility is that there is vestigial ionization left over from a
previous event that heated the gas and collisionally ionized it (Lyu \&
Bruhweiler 1996), such as the passage of a shock front in less a few
times $10^5$yr ago.  If this were the case, it would be difficult to
explain why the gas is not moving at a substantial velocity.  Another
tactic is to propose that most of the gas that we see is rather close to
$\epsilon$ and \bcma, where the hydrogen (and helium) ionization rates
must be much higher than in the local vicinity.

We experimented with ionization models that allowed for the enhancement
of $\Gamma(\hi)$ by placing the clouds closer to \bcma.  For one
possible location, at the nearest point to \acma\ with its white dwarf
companion, the increase in the radiation field is not much: the strength
of the radiation from Sirius~B (Holberg, et al. 1998) reaches a maximum
of only 0.4 times that of \bcma\ as seen near the Sun.  However this
applies only to fluxes observed at wavelengths above the Lyman limit.
If there were
substantially less \hi\ and \hei\ between Sirius~B and the cloud in
front of \bcma, the ionizing fluxes could be larger.

We expect that at a distance of 130~pc from the Sun, a cloud is at the
smallest possible distance from \ecma\ (33~pc) and only 26~pc from
\bcma\ (assuming that the most probable distances between us and the
stars derived from Hipparcos are correct).  Under this condition, the
flux of \ecma\ is enhanced by a factor of 17, and the flux from \bcma\
increases by a factor of 36.  With the spectral distributions given by
Vallerga \& Welsh (1995) for \ecma\ and Aufdenberg, et al. (1999) for
\bcma\ (after correction for absorption by the H and He between these
stars and the Sun), we find that the enhanced fluxes at this position
produce hydrogen neutral fractions of 0.25 and 0.037 for Components C
and D. These values are close to what we observed and reported in
\S\ref{ionization}.

To some degree, our results are dependent on the assumed $N$(\hi) and
$N$(\hei) between the ionizing sources and most of the gas in the
components.  Absorption in front of \bcma\ is constrained by the column
densities that we observed, whereas those for \ecma\ are arbitrary.  For
Component~C our adopted columns were $N(\hi)=1.1~10^{18}$\cmd\ for
\bcma\ and $1.1~10^{17}$\cmd\ for \ecma\ (this low value was required to
raise the ionization high enough).  For Component~D, we made the columns
in front of both \ecma\ and \bcma\ equal to $N(\hi)=2~10^{17}$\cmd.  For
both components, we made the \hei\ to \hi\ ratio in the absorbing gas
consistent with what we found from the ionization equilibrium
calculations.

Our ionization equilibrium calculations indicate that both clouds are
located in a region close to $\epsilon$ and \bcma\ where the radiation
fields are strongly enhanced.  Note also that a physically separate
region, perhaps an outer layer of the cloud representing Component~D that has
evolved to a more fully ionized state, is revealed by the presence of
\siliii.

The fractional ionization of He that we compute is not very substantial. 
Singly ionized He accounts for only 6\% (Component~C) and 19\%
(Component~D) of the total He, even when the material is placed at the
position where the radiation fields from $\epsilon$ and \bcma\ are
enhanced by the large factors given above.  The total column density
$N(\hei)_{\rm C+D}=1.6~10^{18}$\cmd, which is above the lower limits
stated in \S\ref{ionization}.  For Component~C, the computed \hei/\hi\
of 0.38 agrees with the general results of Wolff, et al. (1999), while
the much higher ionization of H in Component~D produces a considerably
larger \hei/\hi=2.2.
\begin{acknowledgements}

The observations reported in this paper are from a guest investigator
program that was approved by NASA as a part of the US share of observing
time on the ORFEUS-SPAS~II flight in 1996, a joint undertaking of the US
and German space agencies, NASA and DARA.  The successful execution of
our observations was the product of efforts over many years by
engineering teams at Princeton University Observatory, Ball Aerospace
Systems Group (the industrial subcontractor for the IMAPS instrument)
and Daimler-Benz Aerospace (the German firm that built the ASTRO-SPAS
spacecraft and conducted mission operations).  Contributions to the
success of IMAPS also came from the generous efforts by many members of
the Optics Branch of the NASA Goddard Space Flight Center (grating
coatings and testing) and from O.~H.~W.~Siegmund and S.~R.~Jelinsky at
the Berkeley Space Sciences Laboratory (deposition of the photocathode
material). 

We are grateful to Martin Lemoine for providing his  absorption line
fitting software that we used to derive the component column densities.
 This research was supported by NASA grant
NAG5$-$616 to Princeton University.
\end{acknowledgements}

\end{document}